\documentclass{Interspeech2024}
\usepackage{multirow}
\usepackage{xcolor}         
\usepackage{tikz}
\usepackage{pifont}
\newcommand{\cmark}{\ding{51}}%
\newcommand{\xmark}{\ding{55}}%

\definecolor{green}{rgb}{0.35, 0.90, 0.63}

\newcommand{\greencheck}{{\cmark}}
\newcommand{\redcross}{{\xmark}}

\interspeechcameraready

\title{LAFMA: A Latent Flow Matching Model for Text-to-Audio Generation}

\name[affiliation={1}]{Wenhao}{Guan}
\name[affiliation={1}]{Kaidi}{Wang}
\name[affiliation={3}]{Wangjin}{Zhou}
\name[affiliation={4}]{Yang}{Wang}
\name[affiliation={4}]{Feng}{Deng}
\name[affiliation={5}]{Hui}{Wang}
\name[affiliation={2}]{\\Lin}{Li$^\ast$}
\name[affiliation={1}]{Qingyang}{Hong$^\ast$}
\name[affiliation={5}]{Yong}{Qin}

\address{
  $^1$School of Informatics, Xiamen University, China\\
  $^2$School of Electronic Science and Engineering, Xiamen University, China \\
  $^3$Graduate School of Informatics, Kyoto University, Japan \\
  $^4$Kuaishou Technology, China  ~
  $^5$Nankai University, China}
\email{\{lilin,qyhong\}@xmu.edu.cn} 


\begin{document}

\maketitle
\renewcommand{\thefootnote}{\fnsymbol{footnote}}
\footnotetext{$^\ast$ Corresponding author.}
\renewcommand{\thefootnote}{\arabic{footnote}}

\begin{abstract}
Recently, the application of diffusion models has facilitated the significant development of speech and audio generation. Nevertheless, the quality of samples generated by diffusion models still needs improvement. And the effectiveness of the method is accompanied by the extensive number of sampling steps, leading to an extended synthesis time necessary for generating high-quality audio. Previous Text-to-Audio (TTA) methods mostly used diffusion models in the latent space for audio generation. In this paper, we explore the integration of the Flow Matching (FM) model into the audio latent space for audio generation. The FM is an alternative simulation-free method that trains continuous normalization flows (CNF) based on regressing vector fields. We demonstrate that our model significantly enhances the quality of generated audio samples, achieving better performance than prior models. Moreover, it reduces the number of inference steps to ten steps almost without sacrificing performance. 
\footnote{ \url{https://github.com/gwh22/LAFMA} .}
\end{abstract}

\section{Introduction}
With the advancements in deep learning and the rapid growth of AI-generated content (AIGC), a lot of content generation methods have emerged across various modalities, including text~\cite{GPT3}, image~\cite{stablediffusion}, video~\cite{sora}, audio~\cite{audioldm,audiogen}, and speech~\cite{valle,mmtts,naturalspeech}. Among these, text-guided audio generation is popular in diverse scenarios such as game sound effects, video dubbing, and virtual reality products.

Existing TTA methods can be broadly categorized into two groups. The first group converts audio into token sequences and uses transformer models for autoregressive generation~\cite{audiogen}. The second group utilizes diffusion models, known for their excellent generation capabilities, to generate audio mel-spectrograms~\cite{audioldm,diffsound}.
DiffSound~\cite{diffsound} leverages a VQ-VAE~\cite{vqvae} model trained on mel-spectrograms to obtain discrete codes. It employs a non-autoregressive token-based diffusion model for audio generation.
AudioLDM~\cite{audioldm} incorporates audio features extracted by a pretrained contrastive text-audio model called CLAP~\cite{clap} during training, and utilizes text features to generate audio during inference. This approach benefits from CLAP's ability to map audio and captions to a shared latent space.
AudioLDM2~\cite{audioldm2} first employs an autoregressive model (AR) to generate AudioMAE~\cite{audiomae} features from text and then uses them to condition the latent diffusion model (LDM). These methods alleviate the reliance on audio-text paired data.
In contrast, Tango~\cite{tango} proposes a different approach, which advocates for instruction-tuned LLMs (Flan-T5)~\cite{flant5} to better comprehend textual descriptions and cross-modal correlations, challenging the concept of embedding audio and text in a shared space.
Auffusion~\cite{auffusion} integrates a powerful pretrained LDM from the Text-to-Image domain to inherit generative strengths and enhance cross-modal alignment.
Audiobox~\cite{audiobox} is a unified model that is capable of generating speech and sound effects based on Voicebox~\cite{voicebox} and SpeechFlow~\cite{speechflow}.

Currently, audio generation models based on latent diffusion models have shown promising results in the field. However, a significant drawback is their reliance on a considerable number of sampling steps to achieve satisfactory performance. While certain methods~\cite{prodiff, comospeech,reflowtts} have implemented acceleration techniques to maintain Text-to-Speech (TTS) performance during fast sampling, these techniques primarily focused on TTS applications. In the case of text-guided audio generation, where the objective is to generate general audios based on holistic text descriptions, the alignment between text and audio is not frame-wise, and the audio information is richer.  Generating high-quality audio becomes more challenging compared to speech generation, especially in a limited number of steps.

In this study, we propose LAFMA, which integrates Flow Matching~\cite{fm} into the audio latent space for text-guided audio generation. Flow Matching is a novel generative method derived from the continuous normalizing flow~\cite{node} framework. It captures the transformation paths that continuously map samples from a simple prior distribution $p_{0}$ to the corresponding samples from the complex data distribution $p_{1}$. 
 Our work is similar to AudioBox, but it  builds the flow matching network on representation of raw waveforms and is fine-tuned upon the SpeechFlow pretraining, which requires multi-stage training.
The contributions of our work are as follows:

\begin{figure*}[t]
  \centering
  \includegraphics[width=0.9\linewidth]{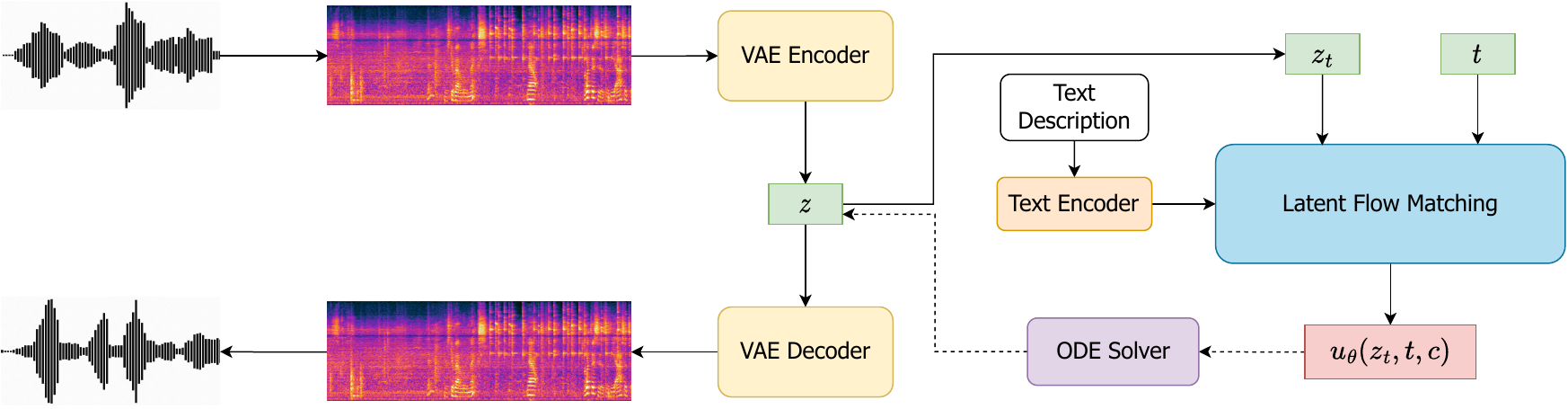}
  \caption{An overview of LAFMA architecture. Note that the dashed line only exists in the inference stage.}
  \label{fig:comp_models}
\end{figure*}

\begin{itemize}
\item We propose LAFMA, a latent flow matching model for text guided audio generation. It can generate high-quality audio samples using a numerical  Ordinary Differential Equation (ODE) solver.
\item We explore the use of classifier-free guidance~\cite{cfg} within the latent flow matching model, leading to improved results in text-conditioned audio generation.
\item Our experiments demonstrate that LAFMA achieves remarkable performance while significantly reducing the number of inference steps required. In particular, we show that LAFMA can generate high-quality audio with only ten inference steps, minimizing the computational burden without sacrificing performance.
\end{itemize}

\section{Flow Matching}
Given the data distribution $p_{1}(x_{1})$ and the Gaussian distribution $p_{0}(x_{0})$, Flow Matching (FM) models the probability path $p_{t}(x_{t})$ directly. This is achieved by considering an Ordinary Differential Equation (ODE) of the form:
\begin{equation}
dx_{t}=v_{t}(x_{t})dt,
\end{equation}
where $v_{t}$ represents the vector field, and $t \in Uniform[0,1]$. The ODE is associated with a probability path $p_{t}(x_{t})$ through the continuity equation. The accurate estimation of the vector field $v_{t}$ by a neural network is sufficient for generating realistic data samples.

The training objective of FM is similar to that of diffusion models~\cite{denoising}. During training, a sample $x_{1}$ is drawn from the data distribution, and a random flow step is sampled. A noisy version of the data $x_{t}$ and its derivative $v_{t}$ for the chosen condition path are computed. A FM model $u_{\theta}$ is trained to predict the derivative $v_{t}$ based on $t$ and $x_{t}$. During inference, to generate a sample $x_{1}$ from the learned data distribution, a sample $x_{0}$ is first drawn from the prior distribution. The ODE solver is then used to estimate $x_{1}$ by integrating the ODE with the derivative parameterized by the FM model.

An alternative approach called optimal transport (OT) was introduced in~\cite{fm}, which utilizes conditional paths with constant directions and speeds. OT paths are easier to learn and can be more accurately estimated by the ODE solver with fewer steps. Empirical evidence from~\cite{fm} demonstrates that the OT path leads to better training and inference efficiency.

For a sample $x_{1}$ and a flow step $t$, the conditional path with OT is given by $x_{t}=(1-(1-\sigma_{\text{min}})t)x_{0}+tx_{1}$ and $v_{t}=x_{1}-(1-\sigma_{\text{min}})x_{0}$, where $x_{0}$ is drawn from the prior distribution $p_{0}(x_{0})$, and $\sigma_{\text{min}}$ is a small value. The objective of FM is defined as follows:
\begin{equation}\label{FM}
\hat{\theta}=\mathop{argmin}_{\theta} E_{t,x_{t}}||u_{\theta}(x_{t},t)-v_{t}||^{2}.
\end{equation}

\section{LAFMA}
LAFMA, as depicted in Figure 1, comprises three key components: i) the text encoder, ii) the latent flow matching model (LFM), and iii) the mel-spectrogram VAE. The text encoder plays a vital role in encoding the input audio description. The resulting encoded textual representation is then utilized to generate a latent representation of the audio by leveraging the flow matching model and an ODE solver. Subsequently, the decoder of the VAE reconstructs a mel-spectrogram based on the latent audio representation. Finally, this mel-spectrogram is passed through a pre-trained vocoder to generate the final audio output.

\vspace{-1mm}
\subsection{Text Encoder}
Similar to Tango~\cite{tango}, we employ the pre-trained LLM FLAN-T5-Large~\cite{flant5} as the text encoder to obtain the text encoding. The FLAN-T5 models are pre-trained on a large-scale chain-of-thought (COT) and instruction-based dataset. This pre-training enables them to effectively leverage in-context information and mimic gradient descent through attention weights, facilitating robust learning of new tasks.

By using FLAN-T5-Large as our text encoder, we benefit from the comprehensive understanding and representation of textual information, which enhances the quality of the encoded text representation. This, in turn, contributes to the generation of accurate and contextually aligned audio outputs in our text-guided audio generation framework.

\begin{table*}[ht!]
\centering
\normalsize
\caption{The comparison between LAFMA and baseline TTA models.  The AS and AC stand for AudioSet and AudiocCaps datasets respectively. $^\clubsuit$ indicates the results are obtained from the original paper. $^\spadesuit$ indicates the results are obtained from the official checkpoints.
The params showed in the table are the trainable params.}
\resizebox{\textwidth}{!}{
\begin{tabular}{cccc|cccc|cc}
\toprule
\multirow{2}{*}{\textbf{Model}} & \multirow{2}{*}{\textbf{Datasets}} & \multirow{2}{*}{\textbf{Text}} & \multirow{2}{*}{\textbf{Params}} & \multicolumn{4}{c|}{\textbf{Objective Metrics}} & \multicolumn{2}{c}{\textbf{Subjective Metrics}} \\ 
& & & & FD~$\downarrow$ & KL~$\downarrow$ & FAD~$\downarrow$ & IS~$\uparrow$ & OVL~$\uparrow$ & REL~$\uparrow$ \\
\midrule
Ground truth & $-$ & $-$ & $-$ & $-$ & $-$ & $-$ & $-$ & $91.83$ & $90.03$ \\
\midrule
DiffSound$^{\spadesuit}$    & AS+AC & \greencheck          & $400$M   & $46.39$ & $2.57$ & $7.81$ & $4.12$ & ${46.70}$ & ${45.67}$ \\
AudioGen$^{\clubsuit}$      & AS+AC+8 others & \greencheck &$285$M  &  $-$    & $2.09$  & $3.13$ & $-$ & $-$ & $-$ \\
AudioLDM-S-Full$^{\spadesuit}$ & AS+AC+2 others      & 
\redcross       & $181$M   & $31.72$  & $2.01$ & $4.11$ & $7.63$ & $69.71$ & $66.33$ \\
AudioLDM-M-Full$^{\spadesuit}$ & AS+AC+2 others       & 
\redcross      & $416$M  & $39.76$  & $2.07$ & $4.91$ & $5.57$ 
 & $73.35$ & $65.12$\\
 AudioLDM-L-Full$^{\spadesuit}$ & AS+AC+2 others & 
\redcross & $739$M  & ${36.35}$   & ${1.85}$ & ${4.90}$ & ${7.51}$ & ${78.63}$ & ${63.69}$ \\
Tango$^{\spadesuit}$ & AC & 
\greencheck & $866$M  & $\mathbf{25.32}$   & $1.45$ & $1.66$ & $7.32$ & $80.16$ & ${75.21}$ \\
\midrule
LAFMA & AC & 
\greencheck & $272$M  & $31.13$   & $\mathbf{1.41}$ & $\mathbf{1.61}$ & $\mathbf{7.71}$ & $\mathbf{80.91}$ & $\mathbf{76.22}$ \\
\bottomrule
\end{tabular}
}
\label{tab:AudioCapsResults}
\end{table*}

\vspace{-1mm}
\subsection{Latent Flow Matching Model for TTA}
Given an input sample $x_{1} \sim p_{1}$, we encode it into the latent representation $z_{1}$ using a pre-trained VAE encoder. In the latent space, our objective is to estimate a probability path that traverses from a random noise $z_{0} \sim p_{0}$ to the source distribution of the latent representation $z_{1}$. To achieve this, we optimize the velocity field network by utilizing the compact dimensionality of the audio latent representations.

We employ the same objective as in the vanilla flow matching, which assumes a constant velocity, while incorporating additional conditional textual information $c$. The optimization objective for the velocity network is expressed as follows:
\begin{equation}\label{LFM}
\hat{\theta}=\mathop{argmin}_{\theta} E_{t,z_{t}}||u_{\theta}(z_{t},t,c)-v_{t}||^{2}.
\end{equation}

To facilitate the sampling process, we utilize an Euler ODE Solver for iterative sampling, which enables us to obtain the predicted latent representation $z$. This predicted representation is then fed into the VAE decoder to generate the mel-spectrogram. Finally, the mel-spectrogram is converted into the audio waveform using a pre-trained vocoder.

\subsection{Classifier-free Guidance}
For diffusion models, the ability to generate controllable outputs can be achieved by incorporating guidance at each sampling step. Taking inspiration from~\cite{guidedflow}, we adapt the classifier-free guidance approach to the conditional vector fields $u_{\theta}(z_{t},t,c)$ in our framework.
During the training process, we introduce randomness by randomly discarding our conditioning information with a fixed probability (10\%). This enables us to train both the conditional LFM $u_{\theta}(z_{t},t,c)$ and the unconditional LFM $u_{\theta}(z_{t},t)$. By incorporating both conditional and unconditional models in the training phase, we ensure that our framework learns to generate high-quality audio outputs under different conditions.

During the generation process, we employ sampling via classifier-free guidance:
\begin{equation}\label{LFM}
\hat{u}_{\theta}(z_{t},t,c)=wu_{\theta}(z_{t},t,c)+(1-w)u_{\theta}(z_{t},t),
\end{equation}
where $w$ denotes the guidance scale. The integration of classifier-free guidance within our framework empowers us to generate diverse and controllable audio samples, providing a trade-off between generational quality and sample diversity.

\subsection{VAE and Vocoder}
The audio vaiational auto-encoder (VAE) compresses the audio  mel-spectrogram $m \in R^{T \times F}$ into an audio representation $z \in R^{C \times T/r \times F/r}$, where $T,F,C,r$ represent the time dimension, the frequency dimension, the number of channels, the compression level, respectively. The encoder and decoder are composed of stacked convolutional modules and are trained by maximizing the evidence lower-bound and minimizing the adversarial loss. In our experiments, $C$ and $r$ are set to 8 and 4, respectively.
We employ HiFi-GAN~\cite{hifigan} as the vocoder to synthesize audio waveforms from mel-spectrograms.

\subsection{Model Configurations}
We freeze the FLAN-T5-Large text encoder in LAFMA and only train the parameters of the LFM model. The flow matching model is based on the Stable Diffusion U-Net architecture~\cite{stablediffusion}. The U-Net has four encoder blocks, a middle block, and four decoder blocks. With a basic channel number of $c$, the channel dimensions of encoder blocks are $[c, 2c,3c,5c]$, where we set $c=128$. We also use a cross-attention dimension of 1024 in the U-Net model. Finally, the number of trainable total parameters is 272M.

\section{Experiments}
\subsection{Experimental Setup}
\subsubsection{Dataset}
We conduct our text-guided audio generation experiments using the AudioCaps~\cite{audiocaps} dataset, which consists of 45,122 audio clips paired with human-written captions for training purposes. The dataset provides a diverse range of audio samples that are ten seconds long, collected from various YouTube videos. We utilize  the AudioCaps test set for evaluation.

The VAE model is trained on a combination of multiple datasets, including AudioSet~\cite{audioset}, AudioCaps, FreeSound, and BBC Sound Effect datasets. All audio clips from these datasets are segmented into ten-second segments and resampled at a frequency of 16KHz.
The LAFMA was trained for 60 epochs using AdamW optimizer~\cite{adamw} with a learning rate of 1e-4 on three NVIDIA 4090 GPUs with a batch size of 8 per GPU. 
\subsubsection{Evaluation Metrics}
We conduct a comprehensive evaluation of our text-guided audio generation system, employing both objective and subjective measures to assess the quality of the generated samples. For objective evaluation, we utilize several metrics including the Frechet Distance (FD), Frechet Audio Distance (FAD), Inception Score (IS), and Kullback-Leibler (KL) divergence.

The FD metric, similar to the Frechet Inception Distance used in image generation, measures the similarity between the generated audio samples and the target samples. FAD, on the other hand, utilizes the VGGish classifier instead of PANNs to calculate the distance, maintaining a similar concept to FD. The IS metric is effective in evaluating both the quality and diversity of the generated samples. Additionally, KL divergence, a reference-dependent metric, quantifies the divergence between the distributions of the original and generated audio samples based on the labels generated by a pre-trained classifier.

In addition to the objective evaluation, we conducted a subjective assessment involving human evaluators. We asked six evaluators to rate 30 randomly selected baseline and LAFMA generated audio samples on two aspects: overall audio quality (OVL) and relevance to the input text (REL). The evaluators provided ratings on a scale from 1 to 100.

\subsubsection{Baselines}
We compare the metrics mentioned above for the samples
generated by the LAFMA and the following systems:
1) GT: This is the ground-truth recording; 2) DiffSound\footnote{\url{https://github.com/yangdongchao/Text-to-sound-Synthesis}}~\cite{diffsound}; 3) AudioGen~\cite{audiogen}; 4) AudioLDM-S-Full~\cite{audioldm}; 
5) AudioLDM-M-Full; 6) AudioLDM-L-Full\footnote{\url{https://zenodo.org/records/7884686}}; 7) Tango\footnote{\url{https://huggingface.co/declare-lab/tango}}~\cite{tango}  .

\begin{table}[ht!]
\centering
\small
\caption{Effect on the objective evaluation metrics with a varying number of classifier-free guidance.}
\resizebox{8cm}{!}{
\begin{tabular}{c|ccccc}
\toprule
\multirow{1}{*}{\textbf{Model}}
& Steps & Guidance & FD~$\downarrow$ & KL~$\downarrow$ & FAD~$\downarrow$ \\
\midrule
\multirow{4}{*}{LAFMA} & \multirow{4}{*}{$25$} 
& $1$ & $47.11$ & $2.32$ & $6.33$ \\
& & $2$ & $38.32$ & $1.46$ & $2.04$ \\
& & $3$  & $\mathbf{37.46}$ & $\mathbf{1.40}$ & $\mathbf{1.85}$  \\
& & $4$ & $37.49$ & $1.41$ & $2.02$ \\
\bottomrule
\end{tabular}
}
\label{tab:Guidance}
\end{table}

\begin{table}[ht!]
\centering
\small
\caption{Effect on the objective evaluation metrics with a varying number of inference steps.}
\resizebox{8cm}{!}{
\begin{tabular}{c|ccccc}
\toprule
\multirow{1}{*}{\textbf{Model}}
& Guidance & Steps & FD~$\downarrow$ & KL~$\downarrow$ & FAD~$\downarrow$  \\
\midrule

\multirow{5}{*}{LAFMA} & \multirow{6}{*}{$3$} & $5$ & $54.89$ & $1.95$ & $9.47$  \\
& & $10$ & $39.68$ & $1.47$ & $3.55$   \\
& & $25$ & $37.46$ & $1.40$ & $1.85$   \\
& & $50$ & $35.96$ & $\mathbf{1.37}$ & $1.68$   \\
& & $100$ & $33.81$ & $1.45$ & $\mathbf{1.48}$  \\
& & $200$ & $\mathbf{31.13}$ & $1.41$ & $1.61$  \\
\bottomrule
\end{tabular}
}
\label{tab:Steps}
\end{table}

\subsection{Results}

\subsubsection{Main Results}
We present the results of our main comparative study in Table \ref{tab:AudioCapsResults}, comparing our proposed method LAFMA with DiffSound~\cite{diffsound}, AudioGen~\cite{audiogen}, Tango~\cite{tango}, and various configurations of AudioLDM~\cite{audioldm}.

In our experiments, we set the sampling steps to 200 for LDM and LFM during inference. And we use a classifier-free guidance scale of 3 for the best results in AudioLDM, Tango and LAFMA.

Notably, LAFMA  surpasses the performance of the AudioLDM-*-Full family models, which were trained on significantly larger datasets  for LDM training. However, compared to Tango, The FD metric is worse. We suppose that it is because of the limitation of model sizes.

Furthermore, LAFMA exhibits highly promising results in subjective evaluation, achieving better overall audio quality score  and  relevance score. These scores indicate its significantly superior audio generation capabilities compared to other text-to-audio generation approaches.

\begin{table}[ht!]
\centering
\small
\caption{Objective metric FD values of AudioLDM-S and LAFMA models with similar parameters at different inference steps when the classifier-free guidance scale is 3.}
\resizebox{8cm}{!}{
\begin{tabular}{cc|cccccc}
\toprule
\multirow{2}{*}{\textbf{Model}} & \multirow{2}{*}{\textbf{Params}}  & \multicolumn{6}{c}{\textbf{Inference Steps}} \\
& & 5 & 10 & 25 & 50 & 100 & 200  \\
\midrule
AudioLDM-S-Full & $181$M & $95.41$ & $78.78$ & $50.50$ & $39.68$ & $34.54$ & $31.72$ \\
LAFMA & $272$M & $54.89$ & $39.68$ & $37.46$ & $35.96$ & $33.81$ & $\textbf{31.13}$ \\
\bottomrule
\end{tabular}
}
\label{tab:compare}
\end{table}

The results presented in our comparative study demonstrate the effectiveness and superiority of our proposed method, LAFMA, in terms of both objective and subjective evaluation metrics. The performance of LAFMA highlights the impact of integrating the LFM for text-to-audio generation, outperforming baseline models  in almost all metrics for text guided audio generation tasks.

\subsubsection{Effect of Classifier-Free Guidance}
We present the impact of varying the guidance scale with a fixed 25 steps in the Table \ref{tab:Guidance}. In the first row, we set the guidance scale to 1, effectively excluding classifier-free guidance during inference. As expected, this configuration performs poorly, significantly lagging behind the classifier-free guided models across all objective measures.
With a guidance scale of 3, we observe the best results in objective metrics including FD, KL, FAD.
These findings highlight the significance of appropriately selecting the guidance scale. A balance needs to be struck in determining the optimal guidance scale to achieve the best results across objective measures.

Our results provide insights into the influence of the guidance scale on the performance of our text-guided audio generation system. By carefully adjusting the guidance scale, we can enhance the quality and fidelity of the generated audio samples, contributing to the overall effectiveness of our approach.

\subsubsection{Effect of Inference Steps}
In our experiment, we utilize Euler ODE Solver for text guided audio sampling. From the changes in various objective metrics including FD, KL, FAD in Table \ref{tab:Steps}, we can observe that as the number of sampling steps increases, the quality of generated audio gradually increases. Moreover, after 10 steps, the trend of sample quality increase is noticeably slow. This also demonstrates the efficient performance of our LFM model in LAFMA, which can result in good sample quality with only ten sampling steps. 
As shown in Table \ref{tab:compare}, we also compare the FD values between our model and AudioLDM-S-Full, which have similar parameters magnitude. We demonstrate that LAFMA outperformed AudioLDM-S-Full at almost all sampling steps by using LFM and FLAN-T5 text encoders, especially for small sampling steps.

\section{Conclusion}
In this paper, we introduced LAFMA, a latent flow matching model for text-guided audio generation, which can leverage a numerical ODE solver to generate high-quality audio samples.
The LAFMA leads to better results for conditional audio generation by employing the classifier-free guidance within the latent flow matching model.
Furthermore, we show that LAFMA achieves remarkable performance with just ten inference steps, effectively minimizing the computational burden without compromising the quality of the generated audio.
The audio samples are publicly available at \url{https://lafma.github.io/}.
\section{Acknowledgements}
Thanks to the National Natural Science Foundation of China (Grant No.62276220 and No.62371407) for funding.

\bibliographystyle{IEEEtran}
\bibliography{mybib}

\end{document}